\documentclass[preprint,12pt]{elsarticle}
\usepackage{amssymb}
\journal{Journal of Molecular Liquids}

\begin{document}

\begin{frontmatter}
\title{The resonant heating of heavy water solutions under the terahertz pulse irradiation}

\author[seu]{Rong-Yao Yang}
\author[gteu]{Zi-Qian Huang}
\author[seu]{Si-Na Wei}
\author[apu]{Qi-Lin Zhang}
\author[seu]{Wei-Zhou Jiang\corref{cor1}}
\ead{wzjiang@seu.edu.cn}

\cortext[cor1]{Corresponding author}
\address[seu]{Department of Physics, Southeast University, Nanjing  211189, China}
\address[gteu]{Department of Physics, Guangxi Teachers Education University, Nanning 530023, China}
\address[apu]{Department of Mathematics and Physics, Anhui Polytechnic University, Wuhu
241000, China}

\begin{abstract}
The heating effect of terahertz pulse with various frequencies and intensities on the heavy water solution  is investigated using the molecular dynamics simulation. Resonant absorptions are found for both heavy water and light water, but at a different resonant frequency which is about 16 THz for heavy water and 21 THz for light water. This resonant phenomenon can be explained perfectly by the collective rotational modes that may release water molecules from hydrogen bonding. The findings not only illustrate the heating mechanism of heavy water solution  under the terahertz pulse irradiation, but also demonstrate a novel difference between light water and heavy water that could have potential applications.
\end{abstract}


\end{frontmatter}


\section{Introduction}
Known as the ``matrix of life", water has attracted tremendous attention since the beginning of human's exploration of the mysterious nature~\cite{franks00}. However, there still exit lots of unrevealed aspects in the dynamics of water,  e.g., see reference~\cite{winkel16}. Among them, one eye-catching topic is the interdisciplinary research of water and terahertz (THz) electric field (TEF)~\cite{burs98,keutsch01,ronne02,heugen06,tiel10,heyden10,mishra13,zhang16,huang16}, partly because THz technology opens a new window to understand the properties of water~\cite{burs98,keutsch01,ronne02,tiel10,heyden10} and also has promising potential applications in biomedicine and imaging~\cite{fitz06,pickwell06,chan07}.
On the other hand, the electric field (EF) heating effect on water has long attained attention. As the applied EF can stem from electromagnetic radiations, one typical application of the EF heating  is the invention of the microwave oven. It has been found that some collective vibrational modes of water are responsible for  the EF heating of water with its efficiency dependent on the EF frequency~\cite{heugen06,heyden10,mishra13,eng03}. For instance, the temperature of water was found to rise from 300 K to 800 K under the irradiation of a 60 ps microwave (0.1 THz) with a root-mean-square intensity of 1 V/nm~\cite{eng03}, while a much more efficient heating was achieved by exposing to a 3 THz pulse with the full width at half maximum less than 1 ps~\cite{mishra13,mishra15}.

Recently, a rotational resonant mechanism was suggested to explain the remarkable enhancement of the water flux through carbon nanotube and the prompt temperature rise of water under the THz radiation exposure~\cite{zhang16,huang16}. Since the principal moments of inertia of water molecules are mainly determined by the hydrogen atoms,  the rotational resonance of heavy water (D$_2$O) should be different for the subrogation of the hydrogen atoms. Heavy water is a very important substance for cooling and energy conduction in nuclear power engineering and has a striking biological effect different from the light water (H$_2$O)~\cite{th60,mo96,ku99}.
Though many features have been explored for light water and heavy water~\cite{burs98,ronne02,mizo92,spchw01,farhi15}, no attention, to our knowledge, has yet been paid to the TEF heating effects on heavy water solution. It is the aim of this work to explore the heating effects on heavy water solution induced by the rotational resonance.  Moreover, the resonant excitation  via the EF absorption can be possibly related to the EF intensity due to the competition of various excitation modes, whereas it has been  seldom investigated. Thus, we will investigate the TEF heating of heavy water solutions  with various TEF frequencies and various intensities as well. In this work, we  use the molecular dynamics simulation to perform the calculations that could provide a vivid picture of the heating process of heavy water solution. We expect to figure out the interaction between TEF and heavy water solution that may reveal the characteristics different from the heating of the simple light water.

\section{Computational Methods}
In this work, the heating process of a water box under the THz pulse exposure is investigated using classic molecular dynamics (MD) simulations. The cubic water box has a length of 2.1 nm containing 289 water molecules. These molecules are randomly the light water or heavy water with the heavy water percentage being 0, 25\%, 50\%, 75\%, 100\%, respectively. The NAMD2 program~\cite{namd05} and CHARMM27 force field~\cite{charmm27} are employed here. We adopt the SPC/E model for light water~\cite{spc87} and SPC/HW model for heavy water~\cite{spchw01}, as these models can achieve a good description for the properties of light water and heavy water, respectively. Periodic boundary condition is performed in all directions. The particle mesh Ewald method~\cite{ewald93} is adopted to handle the electrostatic interaction with a real space cutoff 1.2 nm. The cutoff of van der Waals interaction is also 1.2 nm. The time step is 1 fs, and data are recorded every 20 fs. Preliminarily, a equilibrium state at 300 K is achieved in the simulation using the Langevin temperature control in the canonical ensemble. Using this equilibrium state as an initial state at t=0, we conduct a 10 ps non-equilibrium MD simulation in a uniform THz pulse with the temperature control withdrawn.

Water molecules interact with the THz pulse through the point charge on hydrogen/deuterium atoms and oxygen atoms. The  THz pulse is a Gaussian envelope which is given as
\begin{equation}
E=A\ \cos[2\pi (t-t_c)f] \ \exp[-(t-t_c)^2/(2\sigma ^2)], \label{EF}
\end{equation}
where $A$ denotes the maximum EF amplitude. The  $f,\ t_c$ and 2.355$\sigma$ are the frequency, time center and full width at half maximum (FWHM) of the pulse, respectively. Here, we set A=0.5 V/nm,  $t_c$=5 ps, and FWHM=2 ps. We note that a high power THz pulse with the peak field up to 1 V/nm is achievable nowadays after a progressing development in THz pulse generators~\cite{carr02,vicario14,shalaby15}, and this justifies the usage of A=0.5 V/nm in our simulations. To pinpoint the resonant frequency, we scan the frequency region from 1 THz to 35 THz. The Lorentz force induced by magnetic fields of the THz pulse is neglected here, because it is negligibly small due to the fact that the movement of water molecules is much slower than the speed of light which is also the ratio of $|{\bf E/B}|$ for a electromagnetic wave.

As we know, the temperature fluctuation can not be neglected in a non-equilibrium MD simulation without temperature control. To limit the uncertainty, we define the temperature rise as the difference between the average temperature in the last 3 ps and that in the initial 3 ps. Note that the  hydrogen bond (H-bond) data that include bonds between light water and heavy water are collected in a similar way.   Furthermore, we repeat the simulations six times for each case and then average the results. In so doing, the standard deviation of temperature is suppressed to be less than 5 K.

\section{Results and discussions}
\begin{figure}[thb]
\centering
\includegraphics[height=8.0cm,width=10.0cm]{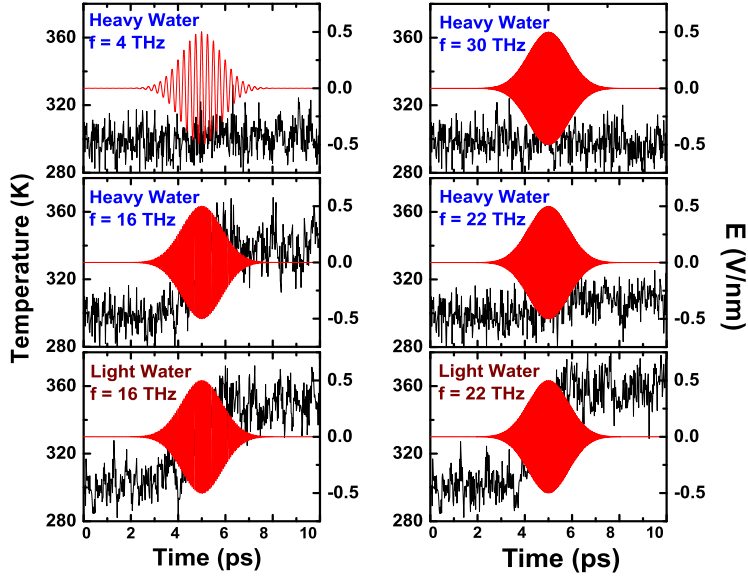}
\caption{(Color online) Temporal evolutions of the temperature and EF   in some representative cases for pure heavy water and pure light water. The left-hand and right-hand vertical axes label the temperature  and EF amplitude, respectively. The top and middle panels represent the cases of pure heavy water with various TEF frequencies, and the bottom panels are for pure light water.} \label{Feg}
\end{figure}

To obtain a straightforward image of the heating process, we illustrate the time evolution of temperature for several representative cases of pure heavy water and pure light water in Fig.~\ref{Feg}, where also shown is the electric field of the imposed THz pulse. From the top panels, we can see that the temperature of the heavy water with the irradiation of the pulse at  $f=4$ or 30 THz does not undergo a noticeable change, fluctuating around 300 K during the whole simulation period. Actually, the situations for light water at these frequencies are quite similar and hence they are not shown here for simplicity. However, the cases with pulses at $f= 16$ and 22 THz are remarkably different. With the pulse at $f=16$  THz, the temperature increases by more than 45 K from 4 ps to 6 ps for both heavy water and light water. With exposure to a 22 THz pulse, the light water gains a temperature rise of more than 60 K, whereas it is much less for heavy water (less than 10 K). These preliminary results indicate that the  water heating by the THz pulse is nontrivial: a difference emerges between light water and heavy water.

\begin{figure}[thb]
\centering
\includegraphics[height=12.0cm,width=8.0cm]{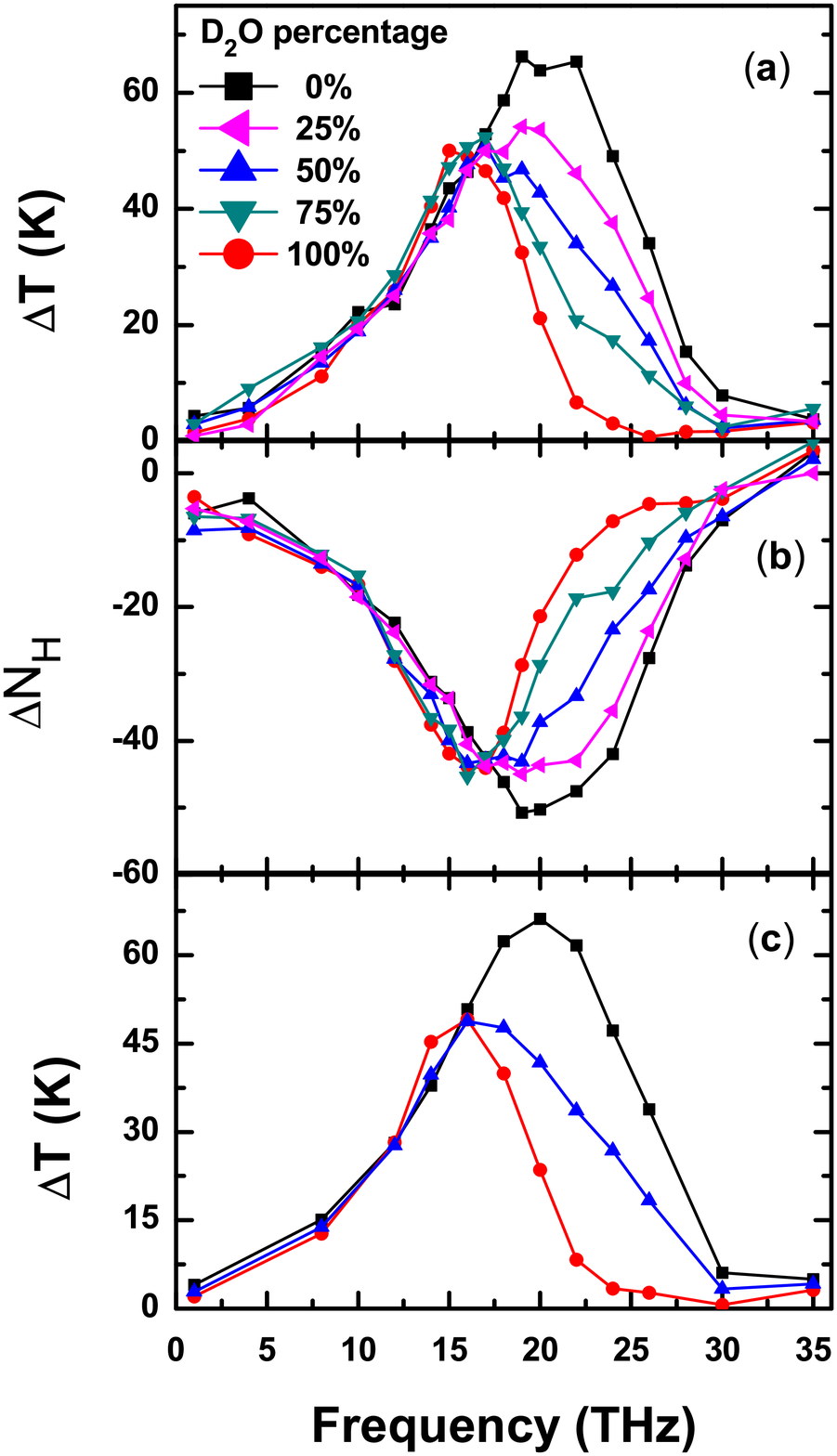}
\caption{(Color online) The temperature rise $\Delta$T (panel a) and the H-bond number variation $\Delta N_H$ (panel b) as a function of the TEF frequency for different heavy water percentages. Panel (c) is the same as panel (a) but for a much larger water box containing 8811 water molecules. Here,  $A$ in Eq.~\ref{EF} is fixed to be 0.5 V/nm. } \label{FTH}
\end{figure}

The sudden rise of the temperature in the aforementioned results suggests that the resonant phenomenon may exist for heavy water and light water in the THz regime. We thus scan the region from 1 THz to 35 THz for different heavy water concentrations. The results of the temperature rise and the corresponding hydrogen bond breaking are presented in Fig.~\ref{FTH}. As can be seen from the top panel,   a peak exists in the frequency profile of the temperature  for all cases. The peak of pure heavy water locates around 16 THz, and the one of pure light water is around 21 THz. The temperature of other cases with various heavy water percentage is peaked at the frequency between 16 THz and 21 THz. Another interesting characteristic shown in the top panel of Fig.~\ref{FTH} is that the peak  broadens noticeably with decreasing the heavy water percentage. In the frequency region of 20-26 THz, a remarkable difference between various heavy water concentration is observed clearly. For example, with a 22 THz pulse, the temperature rises are 6.6, 20.8, 34.0, 46.1, 65.4 K for the heavy water percentage of 100\%, 75\%, 50\%, 25\%, 0, respectively, almost 10 times difference between pure heavy water and pure light water. This indicates that a distinguishable feature between light water and heavy water is amplified significantly under the irradiation of THz pulse at these special frequencies. On the other hand, the tendency below 15 THz or above 30 THz is trivially similar for all cases. As one may concern the dependence of these results on the size of the water box, we show the heating results with a much larger water box containing 8811 water molecules under the same condition in the bottom panel of Fig.~\ref{FTH}. As can be seen, the situation with a larger water box is very similar with that in the top panel. Thus, the size effect in this case is not important.

To understand the resonant behavior, we further analyze the variation in the H-bond that links the molecules in bulk water.  It is known that in bulk water the binding energy per water molecule is mainly located at 15-23 kcal/mol~\cite{hummer01}. If an equal energy is absorbed by water molecules, the H-bond breaking may occur. Displayed in the middle panel of Fig.~\ref{FTH} is the variation of H-bond number due to the THz pulse irradiation. As can be observed from Fig.~\ref{FTH}, the  variation of the H-bond number coincides with that of the temperature rise, albeit opposite with each other. Indeed, the resonant absorption of the THz pulse, responsible for the H-bonds breaking, heats the heavy water solution promptly.

\begin{figure}[htb]
\centering
\includegraphics[height=4.0cm,width=6.0cm]{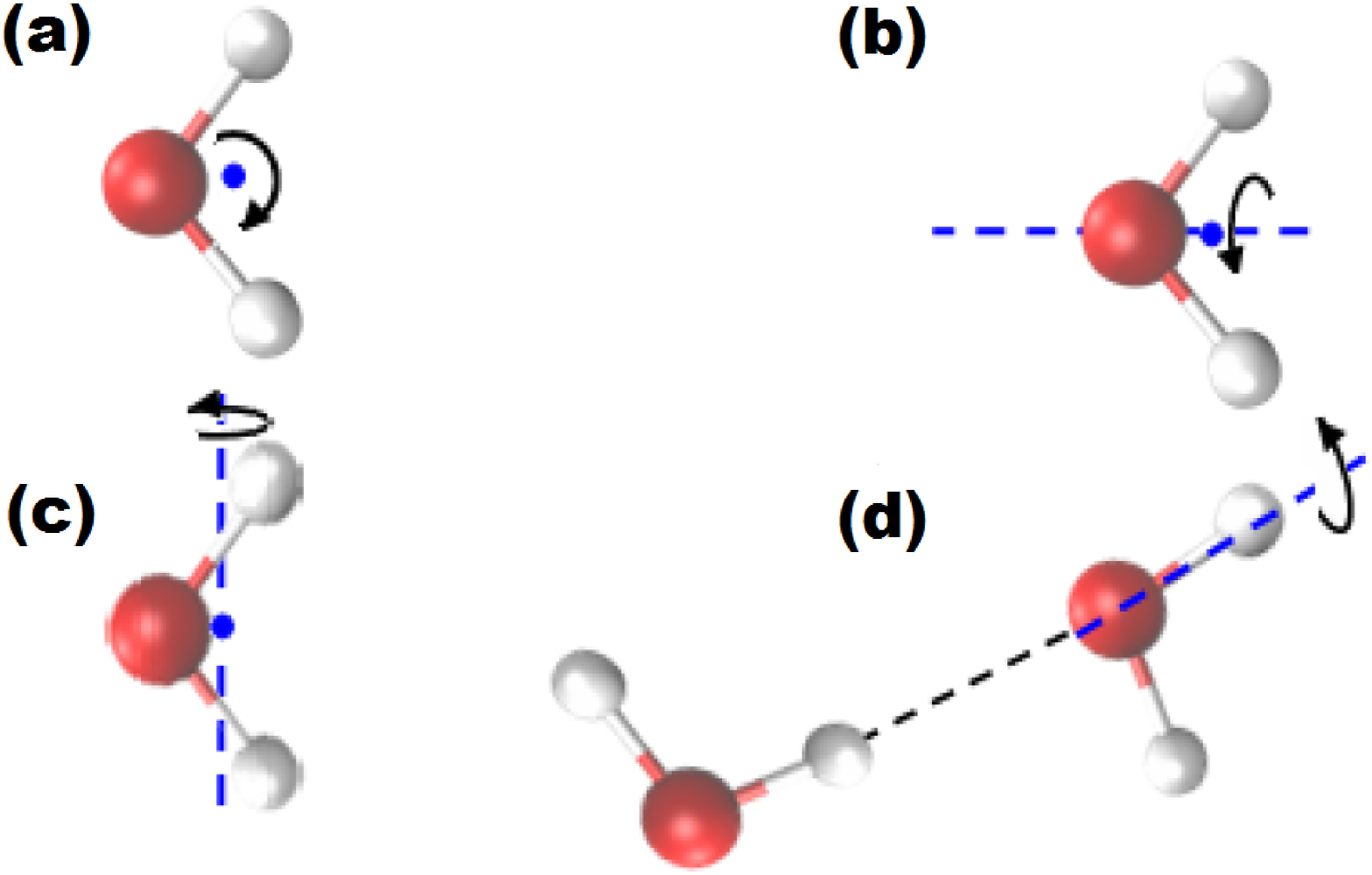}
\caption{(Color online) Several representative librational/rotational modes of water molecules. The blue dots denote the center of mass. The rotational axes in (a), (b), (c) are three principal axes of the water molecule, orthogonal with one another. Note that the axis in (a) is perpendicular to the principal plane. The axis in  (d) is along the direction of the hydrogen bond.} \label{Fmode}
\end{figure}

As already known, the H-bond binding could be broken by the harmonic vibrational mode~\cite{zhang13}, while the resonant frequency  locates  in the range of 5-12 THz for light water and slightly lower for heavy water. The core idea of this vibrational mode is that the energy of water's translational harmonic vibration should be equal to the binding energy of a water molecule. Obviously, the  harmonic vibration can not be used to interpret the resonant structure observed here either for much larger resonant frequency or due to the explicit frequency separation between the light and heavy water. Actually, since the frequency of vibrational resonance is dominated by the mass of oxygen atom, the frequency separation between the light and heavy water is very limited. Nonetheless, the energy transfer through the similar vibrational resonance  of the light and heavy water explains partially why the results of heavy water solution with different concentrations are similar at frequencies below 12 THz.
As the polar water molecules feel the torques in the alternating TEF, the resonant absorption can be aroused by the excitations of librational or rotational modes~\cite{zhang16,huang16}.  We find that the rotor model -- that gives  kinetic energy of these librational or rotational modes of water monomers -- can be used to estimate appropriately the resonant frequencies for both light water and heavy water. More importantly, these modes can be adopted to understand the difference in the heating of the heavy water solution with various heavy water concentrations.
Specifically,  the resonant frequency ($f_c$) of the rotational modes of water molecules is obtained by considering the equality of the binding energy per water molecule ($E_H$) and the kinetic energy of a water molecule librating along a particular axis, i.e. $E_H=\frac{1}{2}I(2\pi f_c)^2$, where $I$ is the moment of inertia corresponding to the particular axis. We mention that the energy of a rotor in this formula should be regarded as the instantaneous kinetic energy of a librational motion, rather than as the energy of a freely rotating water molecule since a free rotation is hindered by the hydrogen-bond networks in this case. We display in Fig.~\ref{Fmode} several representative rotational modes and their revolving axes. With $E_H$=19 kcal/mol for light water~\cite{hummer01} and $E_H$=20.9 kcal/mol for heavy water (the binding of heavy water molecules is 10\% stronger than that of light water~\cite{spchw01}), we can obtain the resonant frequencies of heavy(light) water as 10.9(14.5), 12.9(17.4), 20.4(26.1), and 15.8(21.3) THz, for mode (a), (b), (c), and (d), respectively. Besides these typical cases, if the revolving axes tilt, the estimated frequencies could fall in between those given in Fig.~\ref{Fmode}, yielding a frequency extension of 10.9-20.4 (14.5-26.1) THz for heavy(light) water. The range of the frequency extension is not only remarkably well in accord with the region where the temperature rises promptly and the H-bond number decreases obviously in Fig.~\ref{FTH} for both heavy water and light water, but also consistent with the results calculated from the dipole correlation functions, also see Refs.~\cite{heyden10,chen08}. We  can see that the difference of the rotational resonant frequencies between heavy water and light water is considerable, especially at high frequencies, because the resonant frequencies are related to the moment of inertia of water molecules which is mainly determined by the mass of hydrogen/deuterium atoms. This explains   the explicit frequency separation between the light and heavy water observed in  Fig.~\ref{FTH}. Moreover, we should stress that the present resonant mechanism would be universal for other systems composed of polar molecules  in the TEF.

\begin{figure}[thb]
\centering
\includegraphics[height=6.0cm,width=8.0cm]{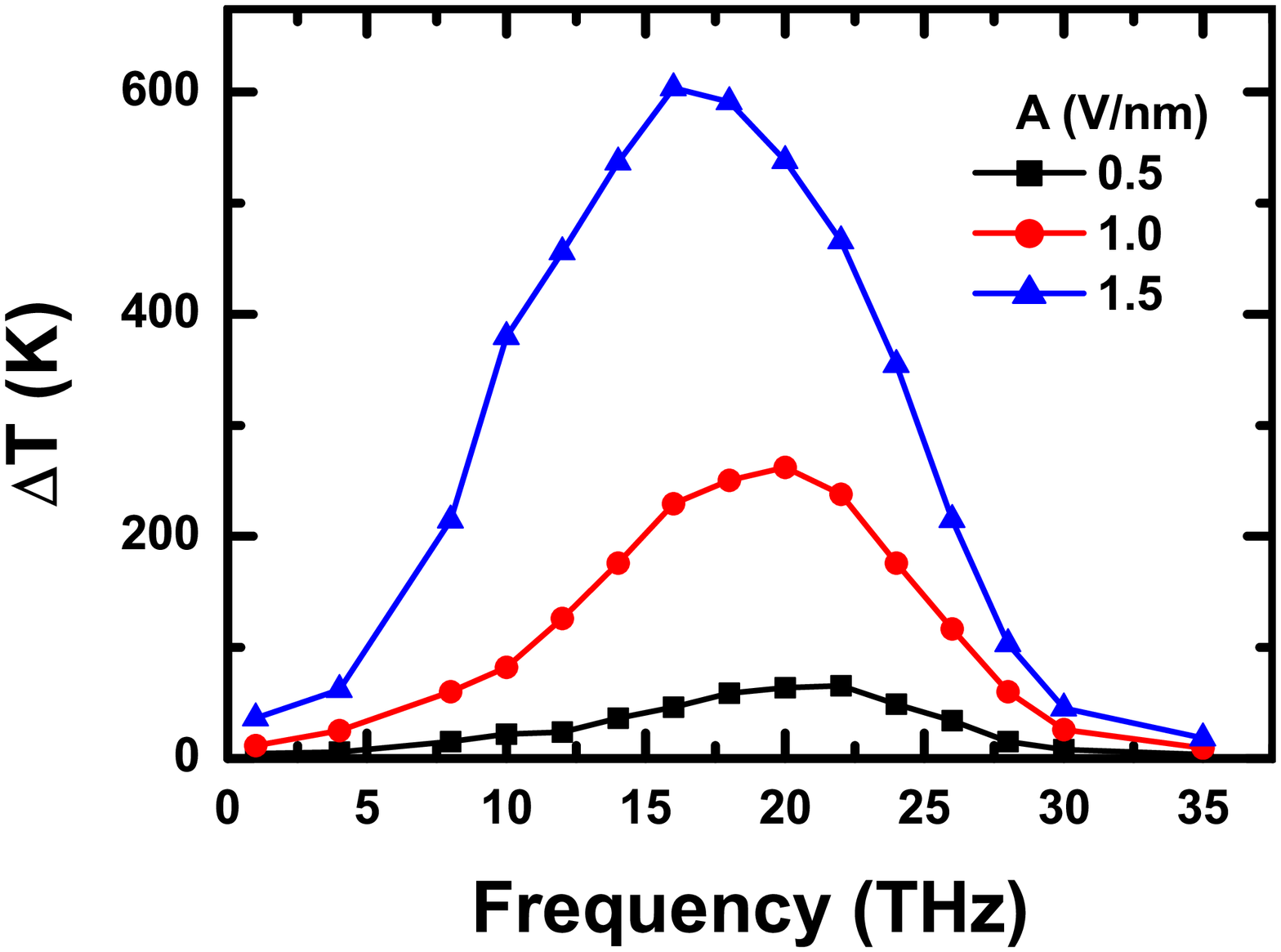}
\caption{(Color online) The temperature rise of pure light water for different TEF amplitude ($A$ in Eq.~\ref{EF}) as a function of the TEF frequency.} \label{FTA}
\end{figure}

The peak structure  in Fig.~\ref{FTH} is a result of the competition and superposition of various modes in a frequency range of  10.9-20.4 (14.5-26.1) THz for heavy (light) water. Due to the different response to the excitation of various modes, the peak can rely on the TEF amplitude.  For instance, the peak for light water heating locates around 19-21 THz with the TEF amplitude of this work, while it could move downward to around 15 THz with a larger TEF amplitude~\cite{huang16}. Evidently, the excitation of the translational resonance is more difficult  as the movement of the center of mass of charge-neutral water molecules can not be induced directly by the EF. For water much cooler than boiling temperature, most water molecules are bound by H-bonds.  In this case, the predominant rotational mode should thus be (d) in Fig.~\ref{Fmode}, corresponding to 15.8 THz for heavy water and 21.3 THz for light water, and this is also the case in Fig.~\ref{FTH}. While the pulse is much more powerful~\cite{huang16}, the efficient heating is not dominated by the (d) mode any more. Specifically, we display  in Fig.~\ref{FTA} the heating curve of pure light water for various TEF amplitudes. We can see explicitly that the peak location shifts downwards to the lower frequency side by increasing the TEF amplitude. This is an interesting phenomenon since a resonance is usually thought to be related to the frequency. In this case, the multiphotonic process~\cite{He08} and the possible change in energy levels with more energy absorbed may also contribute to the shift of the frequency corresponding to the maximum absorption.

In the simulation, we have chosen a Gaussian envelope for  the EF in the THz pulse. If we employ a sinusoidal TEF, the conclusions drawn above remain qualitatively unchanged, because the rotational resonance mechanism, the determinative element, is not altered by the profile of the EF. We have also simulated with a smaller time step ,0.1 fs, and no deviation from the above results has been found. Thus, these collective resonances are quite general and could be used to reveal novel phenomena in other  systems with polar molecules under the TEF irradiation.

\section{Summary}
The heating of various heavy water solutions in the irradiation of THz pulse is investigated with the molecular dynamics simulation. A prompt heating emerges  at the TEF frequency ranging from 16 to 21 THz corresponding to various percentage of heavy water reducing from 100\% to 0. This phenomenon appears with a peak structure resulting from a resonant absorption of the THz pulse due to the excitation of the collective librational/rotational modes. The resonance coincides with the rupture of H-bonds that releases water molecules from hydrogen bonding.  It is  found that the difference in the TEF  heating of heavy water and light water is  manifested as that of the moments of inertia, especially in the frequency region 20-26 THz. In addition, the position of resonant peaks depends moderately on the TEF intensity that ignites various excitation and competition of collective librational/rotational modes.  These findings not only illustrate the heating mechanism of heavy water solutions in the THz pulse, but also demonstrate a novel difference between light water and heavy water that could have potential applications.

\section*{Acknowledgement}
The work was supported in part by the National Natural Science Foundation of China [Grant No. 11275048, No. 11547132, and No. 11604001],  the China Jiangsu Provincial Natural
Science Foundation [Grant No. BK20131286], and the Key Laboratory of New Electric Functional Materials of Guangxi Colleges and Universities [Grant No. DGN201511].


\begin{thebibliography}{99}
\bibitem{franks00} F. Franks, \emph{Water: a matrix of life} (Royal Society of Chemistry, Cambridge, 2000), 2nd Ed.
\bibitem{winkel16} K. Amann-Winkel, R. B\"{o}hmer, F. Fujara, C. Gainaru, B. Geil, and T. Loerting, Rev. Mod. Phys. 88 (2016) 011002.
\bibitem{burs98}B. D. Bursulaya, and H. J. Kim, J. Chem. Phys. 109 (1998) 4911.
\bibitem{keutsch01}F. N. Keutsch, and R. J. Saykally, Proc. Natl. Acad. Sci. U. S. A. 98 (2001) 10533 .
\bibitem{ronne02}C. R\o nne, and S. R. Keiding, J. Mol. Liq. 101 (2002) 199.
\bibitem{heugen06}U. Heugen, G. Schwaab, E. Brundermann, M. Heyden, X. Yu, D. M. Leitner, and M. Havenith, Proc. Natl. Acad. Sci. U. S. A. 103 (2006) 12301.
\bibitem{tiel10}K. J. Tielrooij, N. Garcia-Araez, M. Bonn, and H. J. Bakker, Science 328 (2010) 1006 .
\bibitem{heyden10}M. Heyden, J. Sun, S. Funkner, G. Mathias, H. Forbert, M. Havenith, and D. Marx, Proc. Natl. Acad. Sci. U. S. A. 107 (2010) 12068.
\bibitem{mishra13} P. K. Mishra, O. Vendrell, and R. Santra, Angew. Chem. Int. Ed. 52 (2013) 13685 .
\bibitem{zhang16} Q. L. Zhang, R. Y. Yang, W. Z. Jiang, and Z. Q. Huang, Nanoscale 8 (2016) 1886 .
\bibitem{huang16} Z. Q. Huang, R. Y. Yang, W. Z. Jiang, and Q. L. Zhang, Chin. Phys. Lett. 33 (2016) 013101.


\bibitem{fitz06}A. J. Fitzgerald, V. P. Wallace, M. Jimenez-Linan, L. Bobrow, R. J. Pye, A. D. Purushotham, and D. D. Arnone, Radiology 239 (2006) 533.
\bibitem{pickwell06}E. Pickwell, and V. P. Wallace, J. Phys. D 39 (2006) R301.
\bibitem{chan07}W. L. Chan, J. Deibel, and D. M. Mittleman, Rep. Prog. Phys. 70 (2007) 1325.

\bibitem{eng03}N. J. English, and J. M. D. MacElroy, J. Chem. Phys. 118 (2003) 1589.
\bibitem{mishra15}P. K. Mishra, O. Vendrell, and R. Santra, J. Phys. Chem. B 119 (2015) 8080.

\bibitem{th60}   J. F. Thomson, Ann. N. Y. Acad. Sci. 84 (1960) 736.
\bibitem{mo96} O. B. Mosin,   D. A. Skladnev,  T. A. Egorova, and V. I. Shvets,   Bioorg. Chemis. 22 (1996) 861.
\bibitem{ku99} D. J. Kushner, A. Baker, and T. G. Dunstall,  Can. J. Physiol. Pharmacol. 77 (1999) 79 .

\bibitem{mizo92}K. Mizoguchi, Y. Hori, and Y. Tominaga,  J. Chem. Phys. 97 (1992) 1961-1968.
\bibitem{spchw01} J. R. Grigera, J. Chem. Phys. 114 (2001) 8064.
\bibitem{farhi15}E. Farhi, G. Ferran, W. Haeck, E. Pellegrini, and Y. Calzavara, J. Nucl. Sci. Technol. 52 (2015) 844.
\bibitem{namd05}J. C. Phillips, et al., J. Comput. Chem. 26 (2005) 1781.
\bibitem{charmm27}A. D. MacKerell, et al., J. Phys. Chem. B 102 (1998) 3586.
\bibitem{spc87} H. J. C. Berendsen, J. R. Grigera, and T. P. Straatsma, J. Phys. Chem. 91 (1987) 6269 .
\bibitem{ewald93}T. A. Darden, D. M. York, and L. G. Pedersen, J. Chem. Phys. 98 (1993) 10089.
\bibitem{carr02}G. L. Carr, M. C. Martin, W. R. McKinney, K. Jordan, G. R. Neil, and G. P. Williams, Nature 420 (2002) 153.
\bibitem{vicario14}C. Vicario, B. Monoszlai, and C. P. Hauri, Phys. Rev. Lett. 112 (2014) 213901 .
\bibitem{shalaby15} M. Shalaby, and C. P. Hauri, Nat. Commun. 6 (2015) 5976.
\bibitem{hummer01} G. Hummer, J. C. Rasaiah, and J. P. Noworyta, Nature 414 (2001) 188.
\bibitem{zhang13} Q. L. Zhang, W. Z. Jiang, J. Liu, R. D. Miao, and N. Sheng, Phys. Rev. Lett. 110 (2013) 254501.
\bibitem{chen08} W. Chen, M. Sharma, R. Resta, G. Galli, R. Car, Phys. Rev. B 77 (2008) 245114.
\bibitem{He08} G. S. He, L. S. Tan, Q. Zheng, P. N. Prasad, Chem. Rev. 108 (2008) 1245.
\end{thebibliography}
\end{document}